\documentclass{article}

\usepackage{PRIMEarxiv}

\usepackage[utf8]{inputenc} 
\usepackage[T1]{fontenc}    
\usepackage{hyperref}       
\usepackage{url}            
\usepackage{booktabs}       
\usepackage{amsfonts}       
\usepackage{nicefrac}       
\usepackage{microtype}      
\usepackage{lipsum}
\usepackage{fancyhdr}       
\usepackage{graphicx}       
\graphicspath{{media/}}     

\usepackage{bm}

\usepackage{algorithm}
\usepackage{algpseudocode}

\usepackage{overpic}

\newcommand{\ra}{{\rm a}}

\newcommand{\mvar}[1] {\langle #1 \rangle}

\usepackage{tikz}
\usepackage{pgfplots}
\usepackage{pgfplotstable}
\pgfplotsset{
compat=newest, 
tick label style={font=\footnotesize}, 
}

\title{cPCWE - Perturbed Convective Wave Equation based on Compressible Flows
}

\author{
	\href{https://orcid.org/0000-0002-2148-6703}{\includegraphics[scale=0.06]{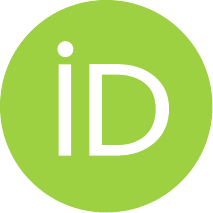}\hspace{1mm}Stefan Schoder} \\
    Institute of Fundamentals and Theory in Electrical Engineering (IGTE)\\
	Graz University of Technology\\
	8010 Graz, Austria \\
	\texttt{stefan.schoder@tugraz.at} \\
}

\begin{document}
\maketitle

\begin{abstract}
This work derives a variant of the perturbed convective wave equation based on the acoustic perturbation equations for compressible flows. In particular, the derivation reformulates the relation of Helmholtz's decomposition to the acoustic and source potential definition. The detailed roadmap of a possible implementation is presented algorithmically. Finally, initial results on the sound prediction capabilities concerning a mixing layer example are presented.
\end{abstract}


\section{Introduction}
\label{sec:Intro} 
Aeroacoustic analogies (e.g., Lighthill's) compute noise radiation efficiently. However, the obtained fluctuating field only converges to the acoustic field in steady flow regions. 
As first recognized by Phillips \cite{PhillipsEQ} and Lilley \cite{LilleyEQ}, the source terms responsible for mean flow-acoustics interactions should be part of the wave operator. This can also be achieved for Lighthill's analogy, which can be adapted for a uniform background flow to account for convection effects inside the wave operator \cite{gloerfelt2003direct}. 

Another approach for computing aeroacoustics is based on a systematic decomposition of the field properties assumed to be related to acoustics and the 'pure' fluid motion. This approach circumvents that sources depend on the acoustic solution and provides a rigorous definition of acoustics inside flow regions (at least for nearly incompressible flows, as shown in \cite{Maurerlehner2022}). Ribner \cite{Ribner1962} formulated the dilatation equation such that the fluctuating pressure is decomposed in a pseudo pressure and an acoustic pressure part $p' = p^0 + p^\mathrm{a} $. Hardin and Pope \cite{HP1994} updated this idea and formulated their viscous/acoustic splitting technique expansion about the incompressible flow (EIF), where they introduced a density correction $\rho_1$. The EIF formalism was modified over the years substantially in \cite{Shen1999,Shen1999a,Slimon1999}.

Being more general, applicable for a wider Mach number range, and starting from the linearized Euler equations (LEE), the field variables $(\rho, \bm u, p)$ are Reynolds decomposed in a temporal mean component
$\mvar{\star}$ and a fluctuating
component~$\star'$. Bailly~\textit{et~al.} \cite{Bailly,Bogey}
indicate important aeroacoustic source terms on the momentum
equation of the LEE. Over the years, the
LEE were modified to guarantee that only acoustic
waves are propagated \cite{Roeck2008}. Significant contributions based on the LEE were derived by \cite{Munz2003,Munz2007,Munz2004,Seo2005,Seo2006}. Ewert and Schröder \cite{Ewert2003} proposed a different decomposition technique, leading to the acoustic perturbation equations (APE). Instead of decomposing the flow field, the source terms of the wave equations are projected onto the acoustic modes obtained from LEE. Hüppe \cite{Hueppe} derived a computationally efficient reformulation of the APE-2 system and named it perturbed convective wave equation (PCWE). Several aeroacoustic low Mach number flow applications have been addressed using the PCWE model successfully~\cite{schoder2019hybrid,schoder2020hybrid,schoder2020computational,schoder2021application,schoder2021aeroacoustic,schoder2019conservative,falk20213d,lasota2021impact,maurerlehner2021efficient,schoder2022opencfs,schoder2022error,tieghi2022machine,schoder2022aeroacoustic,tautz2019aeroacoustic,valavsek2019application,lasota2023anisotropic,schoder2023dataset}. The PCWE is valid for obtaining near-field acoustics of incompressible flows  \cite{Maurerlehner2022}.
Focusing on a two-way coupling of the flow and acoustic variables, Ewert and Kreuzinger developed a computational workflow for low Mach numbers recently \cite{ewert2021hydrodynamic}. A first attempt to establish a scalar wave equation for intermediate subsonic Mach number flows was made by the AWE-PO \cite{schoder2022aeroacoustic,spieser2020modelisation,schoder2022aeroacoustic2,schoder2023acoustic} recently. In this sense, seeking a rigorous definition and an accurate formulation for a large variety of Mach number flows is still a challenge worth investigating. Kempf and Munz \cite{kempf2022zonal} use a high-order DG method for a compressible flow solver with the APE-4 for accurate sound predictions. With the idea of APE and Helmholtz's decomposition, this work aims to extend the PCWE equation for subsonic Mach number flows (the so-called cPCWE).


Regarding the state-of-the-art, we will derive a wave equation based on the fundamental ideas of \cite{Ewert2003}. First, the ideas for incompressible flows are collected and joined to a valid concept (methodological generalization). Second, we extend the concept to compressible flows (physical generalization). The derived aeroacoustic model leads to a stable scalar wave equation that solves an (approximated) compressible potential for subsonic flows. The wave operator only excites longitudinal wave modes (acoustic modes) and includes mean convection effects. The goal of this new wave equation is that it holds for subsonic Mach number flows and recovers the PCWE in the incompressible limit. The investigation on the cPCWE is beneficial to increase further the theoretical understanding of flow-induced sound of subsonic Mach number flows.

\section{Wave equation}

In this section, a convective wave equation based on compressible flow data (cPCWE) is derived. The cPCWE can be derived efficiently from the APE-1 system \cite{Ewert2003} and the use of Helmholtz's decomposition \cite{schoder2019helmholtz}. Firstly, the variables are defined by linearization to distinguish between steady and fluctuating flow components. Secondly, vortical and acoustical perturbations are separated such that the following definitions are obtained
\begin{eqnarray}
p &=&p_0 + p'  = \bar p + p_{\mathrm v} + p_\ra  \\ \rho &=& \rho_0 + \rho'\\
\bm u &=&\bm u_0 + \bm u' =  \bm u_0
+ \bm u_{\mathrm v} + \bm u_{\mathrm a} = \bm u_0 + \nabla \times \bm A - \nabla \psi_\ra \,,
\end{eqnarray}
with the pressure $p$, the mean pressure $p_0$, the perturbation pressure $p'$, the fluid dynamic perturbation pressure $ p_{\mathrm v}$, the acoustic perturbation pressure $p_\ra$, the density $\rho$, the mean density $\rho_0$, the perturbation density $\rho'$, the velocity $\bm u$, the mean velocity $\bm u_0$, the perturbation  velocity $\bm u'$, the fluid dynamic perturbation velocity $\bm u_{\mathrm v} = \nabla \times \bm A$, the acoustic perturbation velocity $\bm u_{\mathrm a} = - \nabla \psi_\ra$ and the vector potential $\bm A$. To eliminate the compressible part of the flow velocity, the Poisson equation
$
    \Delta \, s = \nabla \cdot \bm u' $ has to be solved and the vortical fluctuating velocity can be obtained by $ \bm u_\mathrm{v} = \bm u' - \nabla s \,.
$
We define the acoustic field as irrotational by the acoustic scalar potential $\psi_\ra$. 
Considering a general compressible flow, we arrive at the following perturbation equations
\begin{eqnarray}
\label{eq:APE1}
\frac{\partial p'}{\partial t} + \bm u_0 \cdot \nabla p' +
\rho_0 c_0^2 \nabla \cdot \bm u_\ra &=& 0 \\[2mm]
\rho_0 \frac{\partial \bm u_\ra}{\partial t} + \rho_0 \nabla \big(
\bm u_0 \cdot \bm u_\ra \big) + \nabla p' &=& \rho_0 \nabla \Phi_\mathrm{p}\ \label{eq:momentum}
\end{eqnarray}
with the isentropic speed of sound $c_0$ and the source potential $\Phi_\mathrm{p}$. In this preliminary derivation, we neglect viscous effects and discard the vorticity mode according to \cite{Ewert2003}. 
Rewriting equation (\ref{eq:momentum}) yields the definition of the fluctuating pressure
\begin{equation}
\label{eq:pressAPE1}
p' = \rho_0 \frac{\partial \psi_\ra}{\partial t} +  \rho_0\, \bm u_0\cdot
\nabla \psi_\ra +  \rho_0 \Phi_\mathrm{p} = \rho_0 \frac{D\psi_\ra}{Dt} + \rho_0 \Phi_\mathrm{p} \, .
\end{equation}
The first part accounts for the acoustic pressure $p_\ra$ and the second part is a result of the following Poisson equation \cite{Ewert2003}
\begin{equation}
    \Delta \Phi_\mathrm{p} = - \nabla \cdot  \left[  ( (\bm u_{\mathrm v} \cdot \nabla) \bm u_{\mathrm v})' +    (\bm u_0 \cdot  \nabla ) \bm u_{\mathrm v} +   (\bm u_{\mathrm v} \cdot  \nabla)  \bm u_0   +  T'\nabla s_0   - s'\nabla T_0 \right] \, .
    \label{eq:filter}
\end{equation}
The first part of the source term includes the self-noise of vortical structures, the second and third terms the shear-noise interactions, the fourth and fifth term account for thermal effects. Substituting (\ref{eq:pressAPE1}) into (\ref{eq:APE1}) yields the cPCWE
\begin{equation}
\label{eq:cPCWE}
\frac{1}{c_0^2} \, \, \frac{D^2\psi_\ra}{D t^2} - \Delta \psi_\ra =
- \frac{1}{\rho_0 c_0^2}\, \frac{D \Phi_\mathrm{p}}{D t} \,.
\end{equation}
This convective wave equation describes acoustic sources
generated by compressible flow structures and their wave propagation
through flowing media. In addition, instead of the original unknowns
$p_\ra$ and $\bm v_\ra$, just one scalar
$\psi_\ra$ unknown. As shown in \cite{spieser2020modelisation} and consistent with the pressure correction equation, the fluctuating vortical pressure in the overall domain can be recovered by
\begin{equation}
p_\mathrm{v} = \rho_0 \Phi_\mathrm{p}\,.
\end{equation}
Finally, we have derived a scalar wave equation that separates the source generation processes of compressible flows and the linear acoustic propagation.

\section{Computational workflow}
The algorithm~\ref{alg:cPCWE} can be used to compute the acoustic results. 
From an algorithmic view, we have to do four tasks. Firstly, we are solving the compressible flow equations. Secondly, we solve Poisson's equation to obtain the filtered velocity in every time step
\begin{equation}
    \Delta \, \phi = \nabla \cdot \bm u' \,\,\,\,\,\,\, \mathrm{with} \,\,\,\,\,\,\, \bm u_\mathrm{v} = \bm u' - \nabla \phi \, .
    \label{eq:vortical}
\end{equation}
Every time step can be processed independently. The filtered velocity is used to assemble the cPCWE source. Thirdly, we filter this source with the same operator matrix (\ref{eq:filter}). Finally, we are solving the wave propagation simulation for the time series.
\begin{algorithm}
\caption{Compressible PCWE (cPCWE)}\label{alg:cPCWE}
\begin{algorithmic}
\Require Compressible Flow Simulation
    \State $(\rho, \bm u, p) = \mathrm{fun}(Setup)$
\Require Source computation
    \State $(\bm u_{\mathrm v}) = \mathrm{Laplace-filter}(\bm u')$ using (\ref{eq:vortical})
    \State $(\Phi_\mathrm{p}) = \mathrm{Laplace-filter}(\bm u_{\mathrm 0}, \bm u_{\mathrm v}, s_{\mathrm 0}, s', T', T_{\mathrm 0})$  using (\ref{eq:filter})
    \State $(D_{u_0} \Phi_\mathrm{p}) = \mathrm{fun}(\Phi_\mathrm{p})$ compute the source term field of (\ref{eq:cPCWE}) 
    \State $(f_e) = \mathrm{integrate}(D_{u_0} \Phi_\mathrm{p})$ using the methods in \cite{schoder2021application} to compute the nodal finite element force vector in openCFS-Data\cite{schoder2023opencfs}
\Require Acoustic simulation
    \State $(\rho_\ra, \bm u_\ra, p_\ra) = \mathrm{fun}(f_e, Setup)$, based on the convective wave equation using the finite element method \cite{schoder2022aeroacoustic}
\end{algorithmic}
\end{algorithm}
To sum up, depending on the number of processors and processes one can submit, the algorithm \ref{alg:cPCWE} is of similar complexity than the original PCWE or any other aeroacoustic wave equation. The solution processes for the Laplace equations can be highly parallelized. In the presence of internal wall boundaries near the sources, proper wall boundary conditions must be found for Poisson's equation.

\section{Alternative formulation of the cPCWE}
By inserting the definition of the fluctuating pressure (\ref{eq:pressAPE1}) again into the wave equation (\ref{eq:cPCWE}), a Poisson equation similar to Doak's idea for time-stationary momentum fluctuations \cite{doak1989momentum,schoder2022helmholtz} can be derived
\begin{equation}
\label{eq:cPCWE}
 \rho_0\Delta \psi_\ra =
\frac{1}{c_0^2}\, \frac{D p'}{D t} \,.
\end{equation}

\section{Preliminary results}
As in the previous study~\cite{schoder2023acoustic}, this two-dimensional isothermal mixing layer application is considered to assess the validity of acoustic wave equations. 
The acoustic intensity 
\begin{equation}
  L_\mathrm{I} = 10 \log{ \frac{  I  }{ I_0 } }   
\end{equation}
is used, with $I=\mvar{p'^2}/ ( \rho_{0} c_{0})$ and $I_0=10^{-12} \mathrm{ W.m}^{-2}$ to compare the cPCWE results to the acoustic predictions of Lighthill's equation (LH) and the conservation equations (DNS) as reference. In figure \ref{fig:directivityMM}, the intensity predicted by cPCWE shows good agreement with the one predicted by LH and the DNS. The deviations of the cPCWE and LH results are less than 1.5 dB in the rapid flow region and less than 2 dB in the slow flow region. 

\definecolor{matlabOrange}{rgb}{0.8500, 0.3250, 0.0980}
\definecolor{matlabViolett}{rgb}{0.4940, 0.1840, 0.5560}
\definecolor{matlabGreen}{rgb}{0.4660, 0.6740, 0.1880}
\begin{figure}[ht!]
\centering
\begin{overpic}[scale=0.6,trim=18 0 0 0]{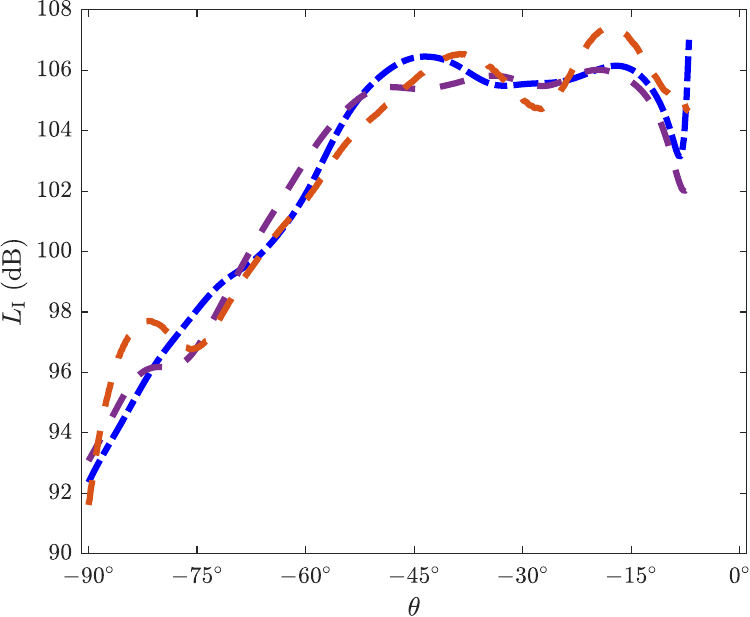}%
\put(10.5,72) {a)}
\end{overpic}
\hspace{0.5cm}%
\begin{overpic}[scale=0.6,trim=18 0 0 0,clip]{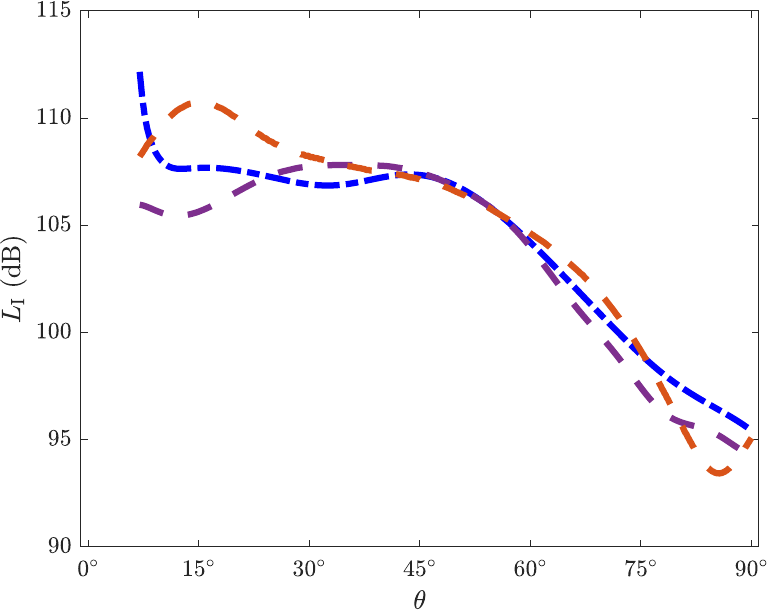}%
\put(10.5,73) {b)}
\end{overpic}
\caption{\label{fig:directivityMM}Preliminary computations of the acoustic intensity $L_{\rm I}$ depending on the angle $\theta$ a) in the rapid flow below the mixing layer and b) in the slow flow region above the mixing layer. 
\textcolor{blue}{-- --}~DNS \cite{schoder2023acoustic}, 
, \textcolor{matlabViolett}{- - -}~Lighthill's equation \cite{schoder2023acoustic}, 
\textcolor{matlabOrange}{- - -}~cPCWE.}
\end{figure}

	\section{Future Work}
	
This short working paper presents the derivation of the cPCWE, and we are happy to receive feedback. First results are presented for a two-dimensional isothermal mixing layer. The convergence of the numerical algorithm is assessed next. Furthermore, we are looking forward to collaborations on the topic to advance the theoretical understanding of flow-induced sound for subsonic flows.

\bibliographystyle{elsarticle-num}
\bibliography{mybibfile}

\end{document}